\documentclass[eqsecnum,aps,ams,preprint,epsf]{revtex4}
\newcommand{\be}{\begin{eqnarray}&&}
\usepackage{graphicx}
\usepackage{wrapfig}
\font\tenbifull=cmmib10 scaled 1200 % bold math italic
\font\tenbimed=cmmib9
\font\tenbismall=cmmib7
\def\bmit{\fam9 }
\mathchardef\bbkappa="7114
\mathchardef\bbrho="711A
\mathchardef\bbsigma="711B
\mathchardef\bbtau="711C
\mathchardef\bbvarrho="7125
\mathchardef\bbvarsigma="7126
\mathchardef\bbPhi="7008
\mathchardef\bbxi="7118

\def\boldtau{{\bmit\bbtau}}

\def\boldPhi{{\bmit\bbPhi}}

\def\bpi{{\mbox{\boldmath$\pi$}}}

\newcommand{\ee}{\end{eqnarray}}

\def\dfrac{\displaystyle\frac}

\begin{document}

\newpage

\thispagestyle{empty}
\title{ $\omega -  \pi\gamma^*$ transition form factor
in proton-proton collisions }

\author{L.P. Kaptari}
\altaffiliation{On leave of absence from
Bogoliubov Lab. Theor. Phys. 141980, JINR,  Dubna, Russia}
\author{ B. K\"ampfer}
\affiliation{Forschungszentrum Rossendorf, PF 510119, 01314 Dresden, Germany}

\date{\today}

\begin{abstract}
Dalitz decays of $\omega$ and $\rho$ mesons, $\omega\to  \pi^0\gamma^*\to\pi^0  e^+e^-$
and $\rho^0\to  \pi^0\gamma^*\to\pi^0  e^+e^-$, produced in
$pp$ collisions are calculated within a covariant effective meson-nucleon theory.
We argue that the $\omega$ transition form factor $F_{\omega\to  \pi^0\gamma^*}$
is experimentally accessible
in a fairly model independent way in the reaction 
$pp \to pp \pi^0 e^+ e^-$ for invariant masses of the $\pi^0 e^+ e^-$
subsystem near the $\omega$ pole.
Numerical results are presented for the intermediate energy
kinematics of envisaged  HADES experiments.
\end{abstract}
\maketitle

\section{Introduction}

The investigation of vector meson production in nucleon-nucleon
($NN$) reactions represents an interesting topic with various implications.
For instance, it is known that the effective repulsive $NN$ forces at
short distances can be described, within a boson exchange model,
by the exchange of $\rho$ and $\omega$ mesons so that a
study of their contribution to the $NN$ elastic amplitude and
to the meson exchange currents in elastic scattering processes off
light nuclei can substantially
augment the knowledge of the short-range part of the $NN$ potential.
Another important issue of vector meson production in $NN$ collisions
is related to electromagnetic probes of strongly interacting systems. As vector
mesons carry the $J^P = 1^-$ quantum numbers as the photon, they couple directly
to real and virtual photons. The latter ones can be converted into di-electrons in
an $s$-channel process, such allowing a direct access to the spectral distribution
of the parent vector meson, even when embedded in strongly interacting matter.
(The strong decay channel products would suffer from final state interaction
with the ambient medium. Thus, the di-electron channel serves as direct or
penetrating probe \cite{Rapp_Wambach}.)

Furthermore, the decay $\omega \to \pi^0 \gamma$ was recently experimentally
studied in photo-excitation of nuclei \cite{CB_TAPS}. The difference of the
strength distribution of the parent $\omega$ for different target nuclei
has been ascribed to a medium modification \cite{our_PRL}.
Such medium modifications are of particular importance for understanding
the electromagnetic emissivities of highly excited, strongly interacting systems,
e.g., created in the course of relativistic heavy-ion collisions.
An extreme option is that the resonances, including the $\rho$ and $\omega$
mesons, are molten once the deconfinement and chirally restored phase is entered
\cite{our_em_papers}.

Another aspect is to supply information on production of vector mesons
in nucleon-nucleon reactions
with similar quantum numbers but rather different quark content, such as
$\omega$ and $\phi$ mesons \cite{ourOmega,ourPhi,nakayama,sibir},
which is interesting with respect to the Okubo-Zweig-Iizuka rule \cite{ozi}
and hidden strangeness in the nucleon.

A particularly interesting subject is the decay of a vector meson.
Besides the above mentioned direct di-electron decay,
$V \to e^+ e^-$, where $V$ stands generically for a vector meson,
valuable information
on the half-off-mass shell decay vertex $V \to \pi  \gamma^*  \to \pi  e^+ e^-$
and related transition form factors (FF) can be obtained.
The functional dependence
of FF's upon the momentum transfer encodes
general characteristics of hadrons, such as charge and magnetic
distributions, size etc. The mentioned $\omega$ transition FF is related to the ratio
of matrix elements 
$\langle \omega \vert \pi^0 \gamma^* \rangle / \langle \omega \vert \pi^0 \gamma \rangle$.
%with normalization $F_{\omega \pi \gamma^*} (0) = 1$.

FF's are also known as important objects for studying bound states within
non-perturbative QCD.
Theoretical tools for  exclusive processes
within non-perturbative QCD are approaches based on light cone sum rules
and factorization theorem (see \cite{rad1,rad,novikov,isgur,isgur1}
and references therein).

In deep-inelastic scattering processes,
an investigation of FF's in a large interval of momentum transfer,
including the time-like region, serves as an
important tool to provide additional information about the various
QCD regimes and on  the interplay between soft and hard
contributions. For instance, it has been found that
the soft part  can be treated as contribution of configurations
in the Fock space  with a minimal number of quark constituents.
This can be considered as a justification for
approaches based on the relativistic quark constituent model for
a covariant treatment of mesons as  two-particle bound states
(see refs.~\cite{moris,anisovich_BS} for details of
covariant description of mesons within Bethe-Salpeter like
approaches); correspondingly computed FF's serve as tests
of models \cite{fred,salme,moris,anisovich_BS,anisovich_IE,anisovich_FF}.

Besides the mentioned QCD-motivated approaches
there is a number of more phenomenological models, e.g., based on
the  dispersion relation technique \cite{anisovich_IE,omegaFormfNormaliz}, or  on
the use of  vector meson dominance (VMD) models \cite{faessler,thomas,meissner}
 or with  effective SU(3) chiral
Lagrangians with inclusion of
the non-Abelian anomaly \cite{meissner,kaiser,nonAbelian}.

Traditionally,  the electromagnetic FF's are studied by electron scattering
off stable particles which provides information  in the space-like region of momenta
where, as well known, the experimental data can be peerless parameterized by dipole formulae.
This in turn  means that in the unphysical region, i.e., for kinematics
unreachable by experiments with on-mass shell particles,
the analytically continued FF's exhibit a pole structure.
Intensively studied FF's
are the ones of the pseudoscalar mesons, chiefly the pion.
Light vector meson FF's have received less attention since their  experimental
determination is more difficult. However, new detector installations, like
the spectrometer HADES \cite{HADES}, can detect di-electrons production in 
proton-proton ($pp$) collisions
in a wide kinematical range of invariant masses with a high efficiency. Thus, a
precision study of the transition FF for the 
$\omega \to \pi^0 e^+e^- $ process becomes feasible.

\begin{figure}[h]  %      Fig1
\includegraphics[width=0.8\textwidth]{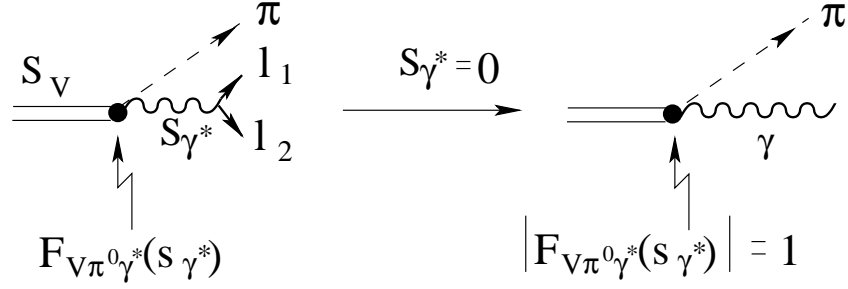} %
\caption{Left part: Dalitz decay of
a vector meson with energy squared $s_V$ into a pion ($\pi$) and a di-electron ($l_1, l_2$).
Right part: the transition form factor
is normalized at the photon point,
i.e.,  $\left| F_{V \pi^0\gamma^* }(S_{\gamma^*=0})\right |=1$.}
\label{fig1}
\end{figure}

The process of vector meson Dalitz decay can be presented as
(see Fig.~\ref{fig1})
\begin{equation}
V\to P +\gamma^* \to P + e^- + e^+
\label{reac}
\end{equation}
where $P$ denotes a pseudoscalar  meson.
Obviously, the probability of emitting a virtual photon is
governed by the dynamical electromagnetic
structure of the "dressed" transition vertex $V\to P$ which
is encoded in the  transition FF's. If the particles
$V$ and $P$ were point like, then calculations of mass distributions
and decay widths  would be straightforward along the standard quantum
electrodynamics (QED) technique. Deviations of the measured quantities from
the QED predictions directly reflect the
effects of  the FF's and thus the internal hadron structure,
and, consequently, can serve as experimental tests  to
discriminate  the different theoretical  approaches.

First experimental measurements of the $\omega$ transition
FF \cite{LandPhysRep,omegaFrmf1,omegaFrmf2,omegaFrmf3} have pointed to a
discrepancy with theoretical pre(post)dictions 
\cite{omegaFormfNormaliz,moris,kaiser}
in the time-like region. Calculations  based on VMD do not
satisfactorily describe the data. A
better description can be achieved with
dispersion relation calculations \cite{omegaFormfNormaliz} or
within models based on the  Dyson-Schwinger equation \cite{moris}.
All these approaches provide rather different  transition
FF's, with the difference increasing with the momentum transfer.
However, the available experimental data is still too scarce
for a preferable choice of the approach, and additional data is needed.
In this context, forthcoming data from the HADES collaboration at the
heavy ion synchrotron SIS18/GSI Darmstadt \cite{HADES}
will substantially contribute to our understanding of the problem.

HADES is a detector installation optimized for studies of processes
with a  $e^+e^-$ pair in one of the
final states in reactions of  hadrons $(p,\pi$)
and various nuclei, i.e., $pp$, $\pi p$, $Dp$, $pA$, $\pi A$, $AA$ etc.\ near the
$\rho$, $\omega$ and $\phi$ thresholds.
In the present paper we study the di-electron production from
Dalitz decay of the lightest vector mesons in $pp$ reactions
at beam energies of a
few GeV for kinematical conditions corresponding to the
HADES setup. Our focus is to investigate
the transition FF $\omega \to \pi^0 e^+e^-$.
To this end we calculate the dependence of the differential cross section
for the reaction $pp \to pp \pi^0 e^+e^-$
upon  the invariant mass of the subsystem $\pi^0 e^+e^-$ around the pole masses
of $\rho$ and $\omega$ mesons and find a kinematical range where
the contribution of $\rho$ is sufficiently small and the cross section
is dominated by Dalitz decays of $\omega$ mesons.
We calculate the double
differential cross section averaged in a suitable kinematical range
as a function of the di-electron invariant mass and argue
that such a quantity, normalized to the real photon point and
supplemented by some specific kinematical factor,
represents the desired transition FF.
In such a way a direct experimental investigation of the 
$\omega$ transition FF is feasible.

Our paper is organized as follows. In section 2 we introduce the $\omega \to \pi^0 \gamma^*$
transition form factor. Section 3 is devoted to the theoretical background for
dealing with the reactions $pp \to pp \omega \to pp \pi^0 e^+ e^-$
and$pp \to pp \rho \to pp \pi^0 e^+ e^-$. Numerical results are
presented in section 4. The conclusions are summarized in section 5, and some
formal relations for an integration procedure are relegated to the appendix.

\section{The transition form factor}

Consider the process of a Dalitz decay of a vector meson into
a pion and a virtual photon (di-electron)  of the type (\ref{reac}).
The effective Lagrangian describing
the vertex $V\to \pi^0\gamma $ reads \cite{meissner,thomas,anisovich_FF}
\begin{equation}
{\cal L}_{ V  \pi^0 \gamma}=f_{ V \pi^0  \gamma}\
\left( \epsilon_{\mu\nu\alpha\beta}
\partial^\mu A^\nu \partial^\alpha \Phi_V^\beta
\right) \boldPhi_\pi^0,
\label{lag1}
\end{equation}
where $A^\nu$ is the electromagnetic four-potential, $\Phi_V$ denotes the neutral
vector meson fields $\omega$ or $\rho$, respectively,
$\boldPhi_\pi^0$ stands for the $\pi^0$ part of the isovector
$\boldPhi_\pi$ pion field,
and $f_{ V \pi^0  \gamma\,}$ is the corresponding coupling constant.
The decay width is calculated from (\ref{lag1}) as
\begin{eqnarray}
\Gamma_{V\to   \pi^0 \gamma}=
\frac{1}{12\pi} \left( \frac{\lambda(s_V,0,\mu_\pi^2)}{4s_V}\right )^{3/2}
f_{ V \pi^0\, \gamma }^2 
\label{g0}
\end{eqnarray}
and serves for a determination of the coupling constant $f_{ V \pi^0\, \gamma}$
from experimental data.
$\lambda$ is the kinematical triangle function,
$\lambda(x,y,z) =(x-(\sqrt{y}-\sqrt{z})^2 )(x-(\sqrt{y}+\sqrt{z})^2)$
and the square of the $\pi^0\gamma$ invariant mass is denoted by $s_V$.
Experimentally,
the branching ratios ${\Gamma_i}/{\Gamma_{tot}}$ for $\omega\to\pi^0\gamma$
and $\rho\to\pi^0 \gamma$ are known,
being $\left (8.9^{+0.27}_{-0.23}\right ) \cdot 10^{-2}$  and
$ \left (6.1\pm 0.8\right ) \cdot 10^{-4}$ \cite{dataGroup}.
Eq.~(\ref{g0}) yields
$f_{\omega   \pi^0 \gamma\,}\simeq 0.72\ \ GeV^{-1}$
and $f_{ \rho  \pi^0 \gamma\,}\simeq 0.25\ \ GeV^{-1}$ for the known total widths
$\Gamma_\omega=( 8.49\pm 0.08)\  MeV$ and $\Gamma_\rho =( 146.5 \pm 1.5)\  MeV$.
For the reaction (\ref{reac}), however, the emitted photon is virtual and,
consequently,  the Lagrangian (\ref{lag1}) must be supplemented by including
the corresponding transition FF  $V\to P$. Then,
for the $\omega$ meson one has
\begin{equation}
\dfrac{d\Gamma}{d s_{\gamma^*}} =  \left( \dfrac{d\Gamma}{d s_{\gamma^*}}\right )_{\rm point}
\left | F_{ \omega  \pi^0 {\gamma^*}\,}(s_{\gamma^*})\right |^2,
\label{defFF}
\end{equation}
where $s_{\gamma^*}$ is the di-electron invariant mass squared.
Formally, eq.~(\ref{defFF}) can be considered as the definition of the transition
form factor. Direct calculation of the diagram Fig.~\ref{fig1}
with the Lagrangian (\ref{lag1}) results in
\begin{eqnarray}
\dfrac{d\Gamma_{\omega\to \pi^0 e^+e^-}}{ds_{\gamma^*}} =
\dfrac{\alpha_{em}}{3\pi s_{\gamma^*}}\frac{\lambda^{3/2}(s_V,s_{\gamma^*},\mu_\pi^2)}{\lambda^{3/2}(s_V,0,\mu_\pi^2)}
\Gamma_{\omega\to  \pi^0 \gamma }  \left | F_{ \omega  \pi^0 \gamma^*\,}(s_{\gamma^*})\right |^2.
\label{dgamma}
\end{eqnarray}
The mass distribution
$\dfrac{d\Gamma_{\omega\to \pi^0 e^+e^-}}{ds_{\gamma^*}}$
is determined by
(i) a purely kinematical (calculable) factor,
(ii) the decay vertex into a real photon (known from experimental data) and
(iii)  the (yet poorly known) transition FF $F_{ \omega  \pi^0 \gamma^*\,}(s_{\gamma^*})$.
Hence, eq.~(\ref{dgamma}) demonstrates that by measuring
the invariant mass distribution  one can get direct experimental access to the 
$\omega$ transition FF \cite{LandPhysRep,omegaFrmf1,omegaFrmf2,omegaFrmf3}.

\begin{figure}[h]  %      Fig2
\includegraphics[width=1.1\textwidth]{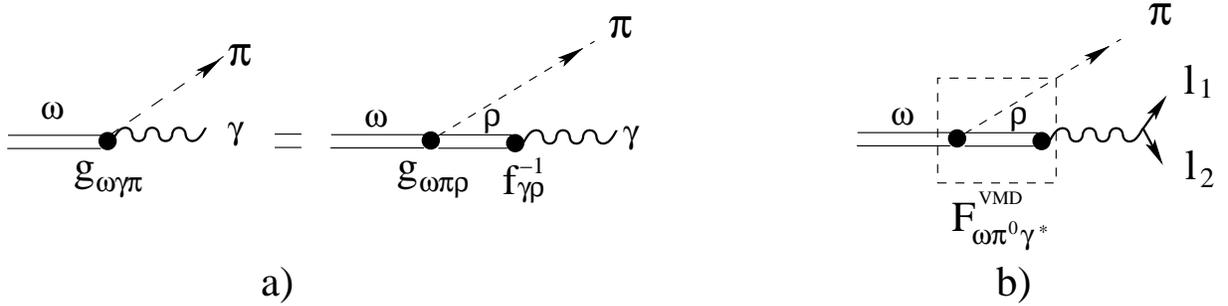} %
\caption{Diagrams for calculation of the transition form factor
$F_{\omega \pi \gamma}(s_{\gamma^*})$ within the VMD model.
Diagrams in a) correspond to  the current-field identity (\ref{currid}),
while diagram b) is the Dalitz decay within the VMD model.}
\label{fig2}
\end{figure}

As mentioned above, the transition FF's are important objects
of theoretical calculations for tests and discrimination  among
the multitude of approaches. The simplest and quite successful
theoretical description of FF's can be performed \cite{faessler,meissner,kaiser}
within the VMD conjecture,
and a reasonably good description of elastic FF's in the time-like region
has been accomplished.
By using the current-field identity \cite{meissner}
\begin{equation}
J^\mu = -e\frac{M_\rho^2}{f_{\gamma\rho}}\Phi_{\rho^0}^\mu  -e\frac{M_\omega^2}{f_{\gamma\omega}}\Phi_{\omega}^\mu
\label{currid}
\end{equation}
with the coupling constants $f_{\gamma\rho}$ and $ f_{\gamma\omega}$
known \cite{friman1,friman2}
from experimentally measured electromagnetic decay widths,
one can  also  compute the transition form
factor  $F^{VMD}_{ \omega  \pi^0 \gamma^*\,}(s_{\gamma^*})$ by
evaluating the corresponding Feynman diagrams (see fig.~\ref{fig2}). 
Contrarily to the elastic case, the FF computed within such an approach 
exhibits disagreement with data (see below).
This immediately implies that with only one (local)
FF it is not possible to satisfactorily describe \cite{kaiser,friman2}
the transition vertex, and the simple $\rho/\omega$ dominance model must be,
at least phenomenologically, supplemented with heavier mesons 
to modify appropriately the shape  of
the transition vertex \cite{faessler}.

\section{The reaction $ \bf pp \to pp \bpi^0  e^+e^-$}
\label{subsecOdin}

\begin{figure}[h]  %      Fig3
\includegraphics[width=0.6\textwidth]{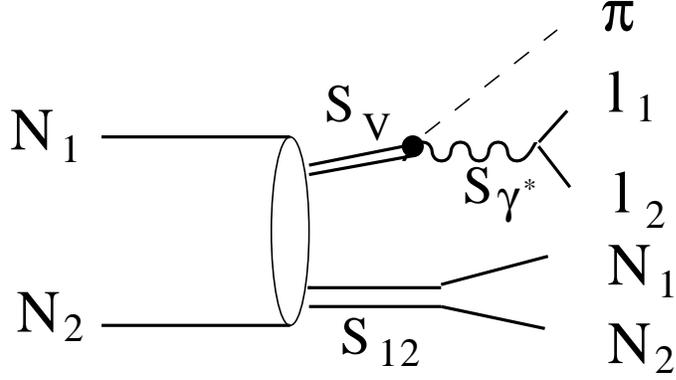} %
\caption{Illustration of the choice of the independent variables
for  the process $NN\to NN + \pi+e^+e^-$ within the duplication
kinematics \cite{bykling}. The invariant mass squared of two final nucleons
is $s_{12}$, while the invariant mass of the subsystem $\pi e^+e^-$  is $s_V$.
The diagram depicted in left part of Fig.~\ref{fig1} enters here as a subprocess.}
\label{fig3}
\end{figure}

Consider now the di-electron ($e^+e^-$) production in the exclusive reaction
\begin{equation}
N_1  + N_2  \to N_1'  + N_2' +\pi^0 + e^+ + e^-  
\end{equation}
for which the process (\ref{reac}) enters as a subreaction.
The invariant cross section is
\begin{eqnarray}
d^{11}\sigma =
\frac{1}{2\sqrt{\lambda(s,m^2,m^2)}}\frac{1}{(2\pi)^{11}}
\frac14 \sum\limits_{\rm spins}\
\,|\ T(P_1',P_2',k_1,k_2,k_\pi,{\rm spins}) \ |^2 d^{11}\tau_f \ \frac{1}{n! },
\label{crossnn}
\end{eqnarray}
where
the factor $1/n!$ accounts for $n$ identical
particles in the final state, $\vert T \vert^2$ denotes the invariant amplitude squared,
and the invariant phase volume $d\tau_f$ is
\begin{eqnarray} d^{11}\tau_f &=&
d s_{12}d s_V  ds_{\gamma^*} \, \label{tau} %\\&&
R_2(P_1+P_2\to P_V +P_{12})
R_2(P_{12}\to P_1'+P_2') \nonumber \\
& \times &
R_2(P_V\to k_\pi +P_\gamma )\
R_2(P_\gamma\to k_1+k_2) 
\end{eqnarray}
with the two-body invariant phase space volume $R_2$  defined as
\begin{equation}
R_2(a+b \to c+d) = d^4 P_c\ d^4P_d\ \delta^{(4)}\left(P_a+P_b-P_c-P_d\right)
\delta\left(P_c^2-m_c^2\right)\ \delta\left(P_d^2-m_d^2\right),
\label{spase}
\end{equation}
where  $P_1, P_2$ and  $P_1', P_2',k_1,k_2,k_\pi$ are the four-momenta 
of the initial and final particles, respectively;
$m$ denotes the nucleon mass, while the electron mass
can be neglected for the present kinematics.
The invariant mass of two particles is hereafter denoted as $s$.
The invariant phase volume $d\tau_f$ in (\ref{tau}) has been chosen
within the so-called "duplication" kinematics \cite{bykling}, i.e. the one which
exploits
invariant two-dimensional phase volumes $R_2$ describing (kinematically)  the
decay of a  real or virtual particle with the invariant mass squared $s$
into two particles, which can also be either real or  virtual.
This kinematics is schematically depicted in Fig.~\ref{fig3}.

The invariant amplitude $T$ is evaluated here
within a phenomenological meson-nucleon  theory based on
effective interaction Lagrangians  which include
scalar ($\sigma$), pseudoscalar ($\pi$),  and
neutral ($\omega$) and charged/neutral vector ($\rho$) mesons (see
\cite{nakayama,nakayama1,ourPhi,ourOmega,ourNuclPhys,titov})
\begin{eqnarray}
{\cal L}_{NN\sigma }&=& g_\sigma \bar N  N \it\Phi_\sigma , \label{mnn1}\\
{\cal L}_{ NN\pi}&=&
-\frac{f_{ NN\pi}}{m_\pi}\bar N\gamma_5\gamma^\mu \partial_\mu
({\boldtau \boldPhi_\pi})N ,\\
{\cal L}_{ NN\rho}&=&
-g_{ NN \rho}\left(\bar N \gamma_\mu{\boldtau}N{\boldPhi_ \rho}^\mu-\frac{\kappa_\rho}{2m}
\bar N\sigma_{\mu\nu}{\boldtau}N\partial^\nu{\boldPhi_\rho}^\mu\right) ,\\
{\cal L}_{ NN\omega}&=&
-g_{  NN \omega}\left(
\bar N \gamma_\mu N {\it\Phi}_{\omega}^\mu-
\frac{\kappa_{\omega}}{2m}
\bar N \sigma_{\mu\nu}  N \partial^\nu \it\Phi_{\omega}^\mu\right),
\label{mnn}
\end{eqnarray}
where $N$ and $\it\Phi$ denote the nucleon and meson fields, respectively,
and bold face letters stand for isovectors.
All couplings with off-mass shell
particles are dressed by monopole form factors
$F_M=\left(\Lambda^2_M-\mu_M^2\right)/\left(\Lambda^2_M-k^2_M\right)$,
where $k^2_M$ is the four-momentum of a virtual particle with mass $\mu_M$.
At $\rho$/$\omega$ threshold-near kinetic energies,
contributions from heavier mesons ($\phi$, $a_1$ ...) can be neglected, and
we consider first only the Dalitz decays of $\rho$ and $\omega$ mesons.

\begin{figure}[h]  %      Fig4
\includegraphics[width=1.0\textwidth]{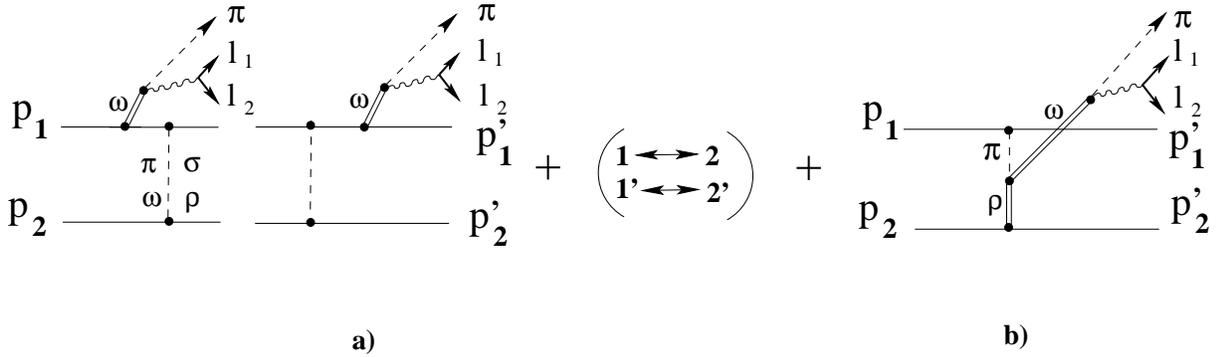} %
\caption{Diagrams for the resonant part of the  process
$N N \to NN + \pi+e^+e^-$ within an effective meson-nucleon theory with
Lagrangians defined by eq.~(\ref{mnn}).
a) Dalitz decay of $\omega$ meson from  bremsstrahlung diagrams,
b) Dalitz decay of $\omega$ mesons from internal conversion.
Analogous diagrams hold for  the $\rho$ meson.}
\label{fig4}
\end{figure}

The Lagrangians (\ref{mnn1} - \ref{mnn}) generate  two classes of Feynman diagrams:
(i) the ones which describe
the Dalitz decay of a vector meson created from nucleon
bremsstrahlung  due to $NN$  interaction (via a
one-boson exchange potential), see Fig.~\ref{fig4}a,
and (ii) Dalitz decay  of a vector  meson, $\omega$ or $\rho$, from a conversion of virtual
$\pi $ and $\rho$ (or $\pi$ and $\omega$)  exchange bosons
into an intermediate  vector meson, i.e., from the internal
$\rho\pi \omega$ vertex, see Fig.~\ref{fig4}b.
The result of a calculation of these diagrams can be cast
in the form of a current-current interaction
\begin{eqnarray}&&
T=J_{\alpha}(12\to 1'2'V)
\left( \dfrac{g^{\alpha\beta}-\frac{P_V^\alpha P_V^{\beta}}{ P_V^2} }{P_V^2 - M_V}
\frac{e\, f_{ V \pi^0 \gamma\,}}{P_\gamma^2}\right)
\left(
\epsilon_{\mu\nu\beta\beta'} P_{\gamma^*}^\mu P_V^{\beta'} j_{\pm}^\nu
\right),
\label{tok}
\end{eqnarray}
where $j_{\pm}^\nu=\bar u(k_1)\gamma^\mu v(k_2)$ is the electromagnetic current
of the final lepton pair, and $J_{\alpha}(12\to 1'2'V)$ stands for the current corresponding
to the vector meson production in $NN$ interaction, i.e.,
the Feynman diagrams $NN\to NNV$ \cite{ourOmega} with the vector meson lines
truncated (cf.\ Fig.~\ref{fig4}).

The amplitude $T$ consists of two parts: 
one ($J_{\alpha}(12\to 1'2'V)$) describing the production of vector
mesons, and the other one $\left(
\epsilon_{\mu\nu\beta\beta'} P_{\gamma^*}^\mu P_V^{\beta'} j_{\pm}^\nu
\right) $ being proportional to the amplitude of Dalitz decays of the
produced mesons. This prominent feature of the amplitude allows to substantially
simplify the expression for the cross section. In the square
of the amplitude one can separate groups of terms  which depend only on
a part of variables  (connected with decay vertices),
and correspondingly the multidimensional integral (\ref{crossnn})
can be partially factorized. Note that the decay part
$\left(
\epsilon_{\mu\nu\beta\beta'} P_\gamma^\mu P_V^{\beta'} j_{\pm}^\nu
\right) $  can also be written in  the form of a current-current interaction
${\cal J}_{(\beta)\nu}(V\to \gamma\pi\, ) j_{\pm}^\nu$.
Note also that all these currents are conserved, i.e. $P_\mu J^\mu=0$.
These circumstances allow one to reduce the dimension of the integral (\ref{crossnn})
by carrying out some of the integrations analytically.
For instance, the summation in the square of the amplitude
over the di-electron spins results in a quantity (known as the leptonic
electromagnetic tensor, see below) which solely contains the whole dependence
upon the momenta of the di-electron. This means that the corresponding integral
over $R_2(\gamma\to l_1+l_2)$ can be evaluated independently of other integrations.
Moreover, since $P_{\gamma *}^2 > 0$ is time-like, one can perform the integration in the system
where the virtual photon is at rest \cite{ourNuclPhys} and
where the integration is particularly simple:
$R_2=d\Omega^*_{\bf k}/8$ and the time components of $l_{\mu\nu}$
vanish. For the leptonic tensor
\begin{equation}
l^{\mu\nu}=4[ k_1^{\mu}P_\gamma^\nu+  k_1^{\nu}P_\gamma^\mu -2 k_1^\mu k_1^\nu -
\dfrac{s_{\gamma^*}}{2}g^{\mu\nu})]
\label{lmunu}
\end{equation}
one has
\begin{equation}
\int l^{\mu\nu}(k_1,k_2,P_{\gamma^*})\, d\Omega^*_{\bf k}=\frac{16\pi}{3} s_{\gamma^*}\left(
-g^{\mu\nu} +\frac{P_{\gamma^*}^\mu P_{\gamma^*}^\nu}{s_{\gamma^*}}
\right).
\end{equation}

In a completely analogous way one can integrate over the phase volume $R_2(V\to \pi\gamma)$
(see Appendix \ref{appA}).
The result is
\begin{eqnarray}&&
\frac{d\sigma}{ds_{\gamma^*} ds_V}=
\frac{1/n!}{2\sqrt{s_0(s_0-4m^2)}}\frac{1}{(2\pi)^{11}}
\frac14
\sum\limits_{\begin{minipage}{0.55cm}
\renewcommand\baselinestretch{0.65} \small\footnotesize nucl.\\ spins \end{minipage}}\,\,\int
|{\cal M}|^2 d^5\tau(s_{12},P_V,P_1',P_2'),
\label{cross1}
\end{eqnarray}
\begin{eqnarray}
|{\cal M}|^2 &\equiv& \int \sum\limits_{spins_\pm} |A|^2 R_2(P_V\to P_{\gamma^*} +k_\pi)\
R_2(P_{\gamma^*}\to k_1+k_2)
\label{matr} \nonumber \\ &=&
\frac{\alpha_{em}}{36}|f_{V\pi\gamma}|^2 \frac{(2\pi)^3}{s_V s_{\gamma^*}}
\lambda^{\frac32}(s_V,s_{\gamma^*},\mu_\pi^2)\nonumber \\ & \times &
J_\alpha(NN\to NNV) 
\left( \displaystyle\frac{-g^{\alpha\beta} + \frac{P_V^\alpha P_V^\beta}{P_V^2}}{P_V^2 -M_V^2}
\right) J_\beta^+(NN\to NNV),
\end{eqnarray}
where the phase volume corresponding to the process of pure
vector mesons production in $NN$ interaction, and
$ds_{12} R_2(P_1+P_2\to P_V + P_{12})  R_2( P_{12}\to P_1'+P_2')$
is denoted as
$d^5\tau(s_{12},P_V,P_1',P_2')$. In principle, since $P_V^\alpha J_\alpha=0$,
the term proportional to $\frac{P_V^\alpha P_V^\beta}{P_V^2}$ can be omitted. 
We keep it for further convenience for the interpretation of results.

\section{Results}

Expressions (\ref{matr}) and (\ref{cross1}) determine the cross section for di-electron
production within the effective meson-nucleon theory.
In our calculations of the nucleonic current $ J_\alpha(NN\to NNV)$
we use  the explicit expressions for the conversion and bremsstrahlung diagrams
quoted in ref.~\cite{ourOmega}. As mentioned above, the Dalitz decay of the $\rho$ meson
also contributes as interference effect,
so that the  current
$ J_\alpha(NN\to NNV)$ and, consequently, the total amplitude $T$ is
a sum of two terms.
Since both $\rho$ and $\omega$ are not stable the corresponding masses
receive  imaginary parts, i.e., $M_V\to M_V -iM_V \Gamma_V/2$,
where $\Gamma_V$ is the total decay width of the  respective vector meson.
The  $\rho$ meson decays mainly into two pions. Consequently,
its width, as a function of the invariant mass $s_V$ is given by
\be
\Gamma_\rho (s_V)=\Gamma_\rho(M_\rho^2)\frac{M_\rho^2}{s_V}\left(
\displaystyle\frac{\sqrt{s_V-4\mu_\pi^2}}{\sqrt{M_\rho^2-4\mu_\pi^2}}\right)^{3},
\ee
where  $\Gamma_\rho(s_V=M_\rho^2) \approx 0.15\  GeV$. The width of the $\omega$ meson
has been kept constant $\Gamma_\omega \approx 0.009\ GeV$ in
the present calculations. All
other effective constants entering into the Lagrangians
(cut-off  form factors, coupling constants, meson masses) have been taken from
ref.~\cite{ourOmega}.
Final state interaction (FSI) among the nucleons have been calculated within
the Jost function formalism \cite{gillespe}
which reproduces the singlet and triplet phase shifts
at low energies. In principle,
the nearly on-mass shell $\omega$ and $\rho$ mesons in the intermediate states can also
interact with the nucleons. The magnitude of such corrections
has been estimated in ref.~\cite{omegaFSI}
by a simulation of rescattering vector mesons off nucleons.
%with  phenomenological value   of the  cross sections
%($\sigma_{\omega N}\sim 40 mb$).
The result is that FSI effects from the $\omega$ meson   rescattering
are small. Consequently, due to the finite life time of the $\omega$ meson,
the reaction product from the Dalitz decay, the pion, is separated in
time-space from the nucleons, and effects of $\pi N$
rescattering in the final state have not been included.

\begin{figure}[h]  %      Fig5
\includegraphics[width=0.8\textwidth]{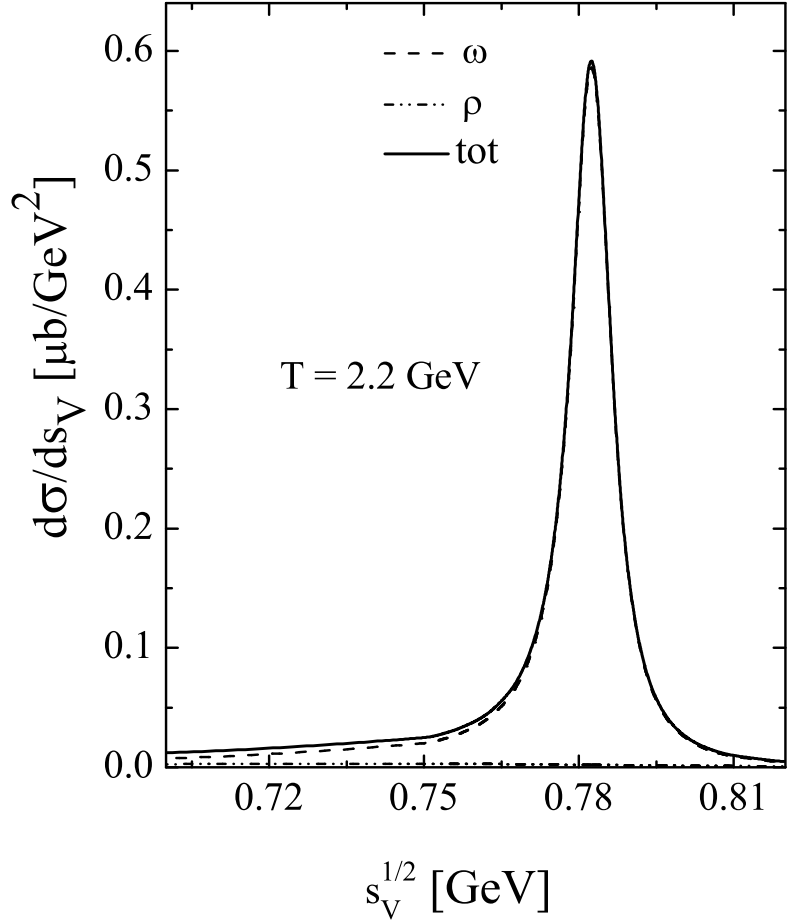} %
\caption{Differential cross section  $d\sigma/ds_V$ for the reaction
$pp\to pp + \pi+e^+e^-$ for  the kinetic beam energy $T_{beam}=2.2\ GeV$.
$s_V^{1/2}$ is the invariant mass of the subsystem
$\pi e^+e^-$. Dashed and dot-dashed lines are contributions from decay of
$\omega$ and $\rho$ mesons,
respectively, while  the dotted line is the total distribution with interference
effects. The transition form factors, entering as input into
the calculations, have been computed within the VMD Model
(see Fig.~\ref{fig2}). }
\label{fig5}
\end{figure}

\begin{figure}[h]  %      Fig6
\includegraphics[width=0.8\textwidth]{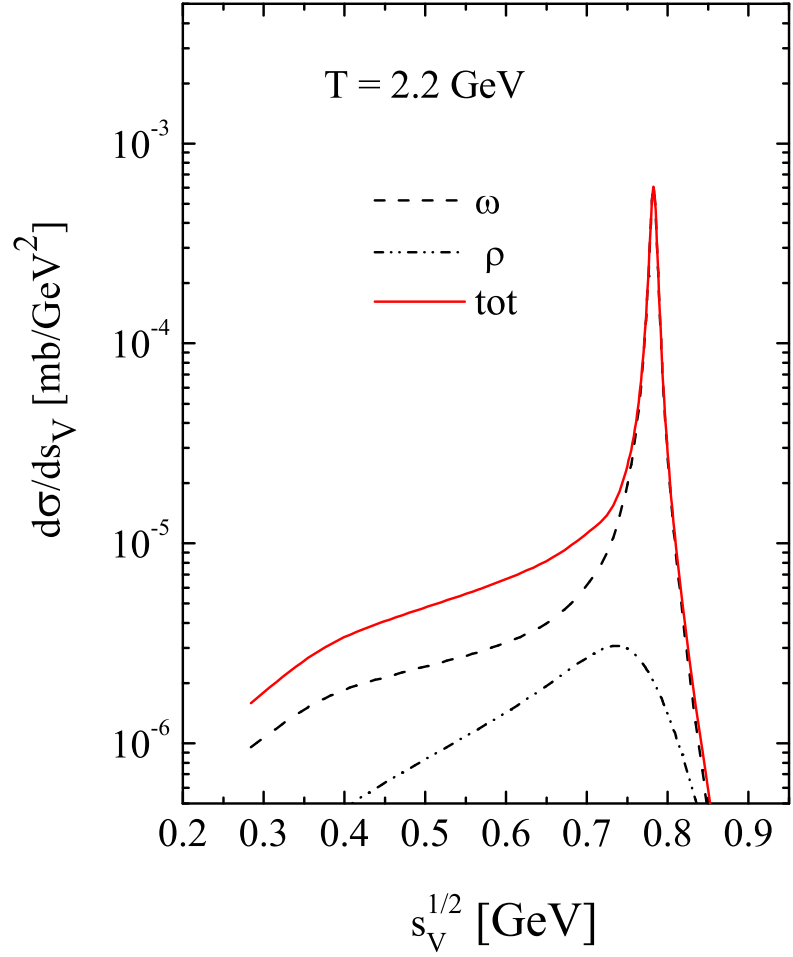} %
\caption{The same as in Fig. \ref{fig5} but in a log scale.
Left to the  $\omega$ peak the contribution of the $\rho$ meson manifests itself
as interference effect.}
\label{fig6}
\end{figure}

Results of calculations of the mass distribution  $d\sigma/ds_V$
are presented in Figs.~\ref{fig5} and \ref{fig6} in linear and log scales, respectively.
We have chosen as  kinetic beam energy  $T_{beam}=2.2\, GeV$, similar to the HADES
proposal \cite{HADES}.
The  dashed line is the contribution from Dalitz decay of  $\omega$
mesons $\omega\to \pi^0 e^+e^-$,  the dot-dashed
line is the corresponding  $\rho$ meson contribution, and the solid curve
is the total cross section, including interference effects as well.
It can be seen that in the very vicinity of the $\omega$ pole the contribution from
$\rho$ mesons is fairly
small. This is an understandable result, since the branching ratio
for the Dalitz decay of $\rho$ meson is much smaller than that for the
$\omega$ meson \cite{dataGroup}. However, as seen from Fig.~\ref{fig6},
outside the $\omega$ pole mass the interference effects are rather significant.
Note that in the direct di-electron bremsstrahlung
(two-body channel decay of vector mesons)
the contribution of $\rho$ can be competitive with that of $\omega$ \cite{ourNuclPhys}.

The obtained results in Figs.~\ref{fig5} and \ref{fig6} persuade us that
for the invariant mass of the $\pi e^+e^-$ subsystem close to the $\omega$ pole
mass, the contribution from $\rho$ can be disregarded. This
also implies that in the double differential cross section
there is a suitable interval in vector meson mass $s_V$ in which the contribution
from $\rho$ can be neglected.

\begin{figure}[h]  %      Fig7
\includegraphics[width=0.8\textwidth]{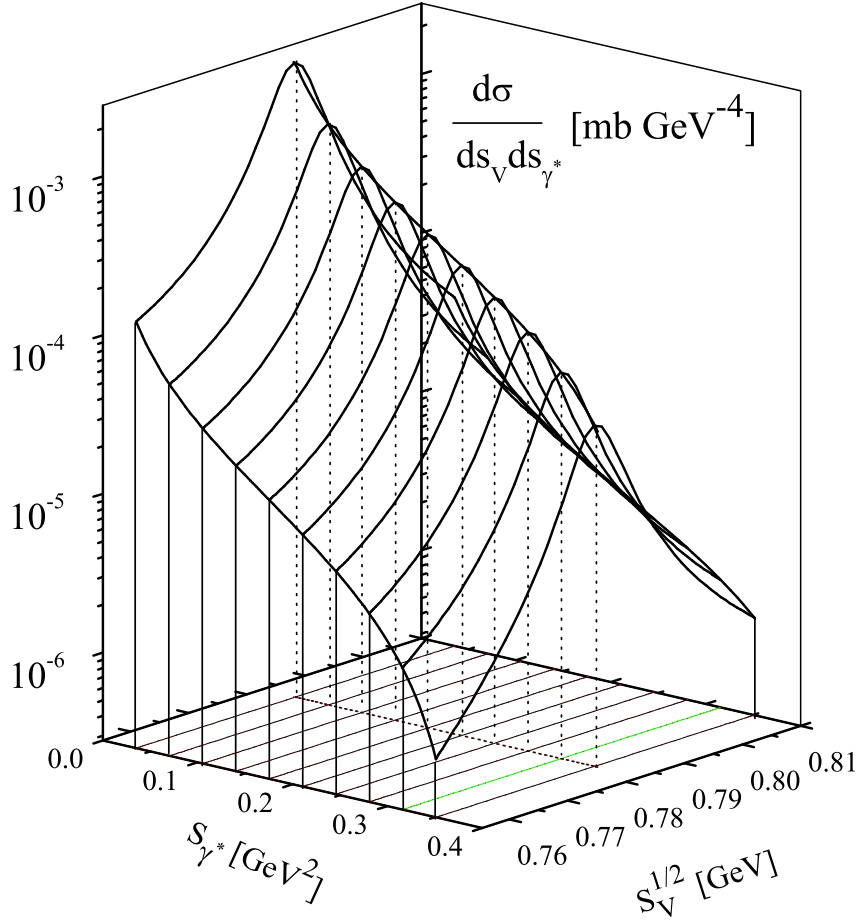} %
\caption{ The double differential cross section
$d\sigma/ds_Vds_{\gamma^*}$  in the  vicinity of the $\omega$ pole
mass $s_\omega^{1/2}=0.782 GeV$ as
a function of the di-electron invariant mass squared $s_{\gamma^*}$. The
interval for the  $\omega$ mass has been chosen with respect   to
the envisaged HADES resolution $\sim  3.5\%$ \cite{HADES}.}
\label{fig7}
\end{figure}

In Fig.~\ref{fig7},  results of calculations of the  double differential cross section
$d\sigma/ds_Vds_{\gamma^*}$ are presented as a function of the invariant mass squared
of the di-electron, $s_{\gamma^*}$, in a narrow bin covering
the $\omega$ meson pole, i.e.\ at mass $s_V\sim M_\omega^2$.
It can be seen that in the whole kinematical range
of the di-electron invariant mass  the double differential cross section
$d\sigma/ds_Vds_{\gamma^*}$
displays a narrow pronounced peak, which is governed by
contributions from Dalitz decays of $\omega$ mesons. This means that by
selecting events with invariant masses
$s_V$ of the $e^+e^- \pi$ system in this interval and varying the invariant mass
$s_{\gamma^*}$ of di-electrons, one can experimentally study the process (\ref{reac})
in $pp$ collisions. Let us recall in this context the studies 
\cite{friman1,rho_omega_interference},
where for the exclusive reaction $\pi N \to N e^+ e^-$ the quantum interference
of intermediate $\rho$ and $\omega$ mesons have been analyzed. 
In certain kinematical regions
this interference is fairly severe and may hamper a clear distingtion
of $\rho \to e^+ e^-$ and $\omega \to e^+ e^-$ contributions.
In this respect, our calculations support a good prospect to isolate the
$\omega \to \pi e^+ e^-$ subreaction vs.\ the $\rho \to \pi e^+ e^-$ part
in the exclusive reaction $pp \to pp \pi^0 e^+ e^-$.
For a further discussion of interference effects see below.

Let us now focus on the part of the diagrams describing the Dalitz decay of the
produced  vector mesons in the vicinity of the invariant mass $s_V=s_\omega$.
If the contribution from only one vector meson
(e.g., the $\omega$ meson) is taken into account then, as
seen from eqs.~(\ref{matr}) and (\ref{cross1}), the cross section can be
presented in the factorized form
\begin{eqnarray}&&
\frac{d^2\sigma}{ds_{\gamma^*} ds_V}= \frac{\sqrt{s_V}/\pi}{\left (
s_V -M_V^2\right)^2}\, \int d^5\sigma^{tot}\left (NN\to NNV
\right)\, \frac{d\Gamma(V\to \pi^0e^+e^-)}{ds_V}, \label{twostep}
\end{eqnarray}
where
\begin{equation}
d^5 \sigma^{tot}\left (NN\to NNV\right)\,=
\frac{1}{2\sqrt{\lambda(s_0,m^2,m^2)}}\frac{1}{(2\pi)^5}\frac14
 \sum\limits_{\begin{minipage}{0.55cm}
\renewcommand\baselinestretch{0.65} \small\footnotesize nucl.\\ spins \end{minipage}}\, \sum_{\lambda_V}
|(J \xi_{\lambda_V})|^2
d^5\tau(s_{12},P_V,P_1',P_2')
\label{twostep1}
\end{equation}
is exactly the total cross section of  real vector meson production
in $NN$ reactions \cite{ourOmega,ourNuclPhys}.
In eq.~(\ref{twostep1}) we  formally   introduced a polarization vector $\xi_{\lambda_V}$
which corresponds to a real vector meson $V$ with  mass $s_V$,
$\sum_{\lambda_V}\xi_{\lambda_V}^\alpha \xi_{\lambda_V}^{+\beta}= -g^{\alpha\beta} + \frac{P_V^\alpha P_V^\beta}{P_V^2}$.
Note that eqs.~(\ref{twostep}) and (\ref{twostep1})
can be easily generalized for contributions from few mesons:
in such a case, the cross section will consist on a sum of two-step like
cross sections (\ref{twostep}), corresponding to each meson, and interference terms.
In the mentioned kinematical bin our cross section
coincides with the one obtained within a two-step mechanism with one isolated meson.
However, outside this kinematical region this is not longer the case, since, apart from
interference effects, even the cross section (\ref{twostep1}) is not anymore
an experimentally well defined quantity, but rather describes the  production of
a (deeply) virtual vector meson $V$
(see discussion in \cite{ourNuclPhys}).

From (\ref{dgamma}), (\ref{twostep}) and (\ref{twostep1}) it can be seen that
the dependence upon the kinematical variables of the subprocesses $NN\to NNV$
and $\omega\to \pi^0 e^+e^-$ can be, in principle, separated in a model independent way by
performing measurements of the double differential cross section $d^2\sigma/ds_{\gamma^*} ds_V$
keeping the invariant mass $s_V$ constant and varying the di-electron mass $s_{\gamma^*}$.
In such a way one can extract the transition FF in  the same
manner as in \cite{LandPhysRep,omegaFrmf1,omegaFrmf2,omegaFrmf3}: Define the quantity
\begin{equation}
\left |F(s_{\gamma^*})\right |^2=
\dfrac{s_{\gamma^*}}{s_{min}}\frac{\lambda^{3/2}(s_\omega,s_{min},\mu_\pi^2)}{\lambda^{3/2}(s_\omega,s_{\gamma^*},\mu_\pi^2)}
\frac{\left \langle {d^2\sigma}/{ds_{\gamma^*} ds_V}\right \rangle}
{ \left \langle{d^2\sigma}/{ds_{\gamma^*} ds_V}\right  \rangle  |_{s_{\gamma^*}=s_{min}}},
\label{theorformf}
\end{equation}
where $\langle\ldots \rangle$ denotes an average about the $\omega$ pole mass corresponding
to the experimental mass resolution
(say, $3.5$ \% as envisaged for forthcoming measurements at HADES \cite{HADES}), and $s_{min}> 4m_e^2$ is
the minimum value of the di-electron mass which plays a role of a normalization point.
Then, as seen from eqs.~(\ref{dgamma}) and
(\ref{twostep}), in the kinematical range, where the contribution of Dalitz decays
of $\rho$ mesons and  interference corrections are negligible,
the defined quantity (\ref{theorformf}) represents indeed the transition
FF $F_{\omega \pi^0 \gamma^*}$.

\begin{figure}[h]  %      fig8
\includegraphics[width=0.8\textwidth]{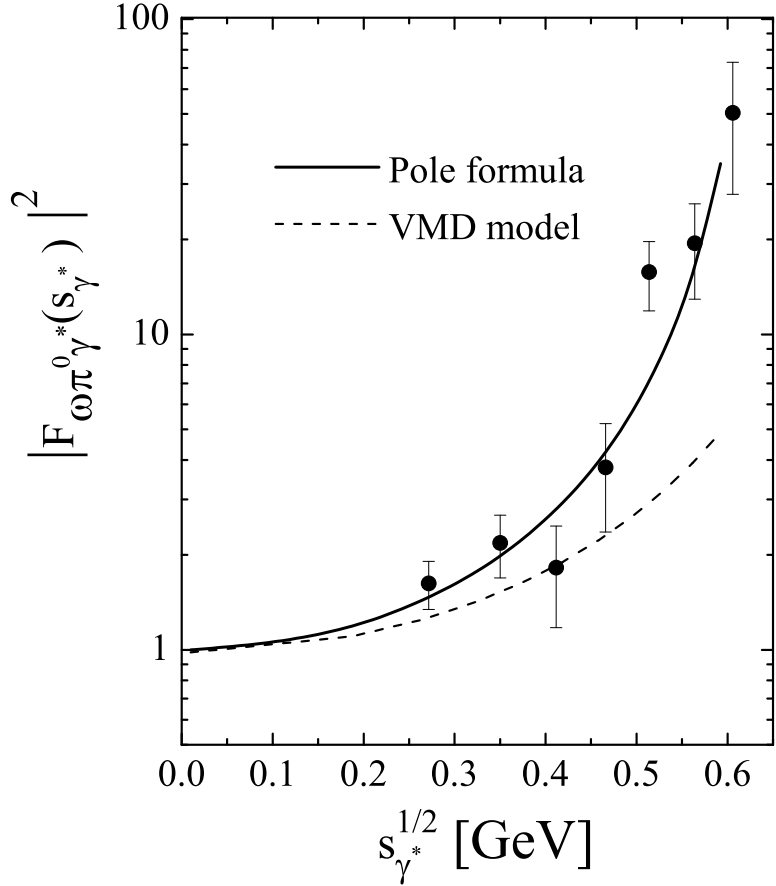} %
\caption{The ratio (\ref{theorformf}) calculated at $T_{beam}=2.2\, GeV$
with two different transition
form factors: the dashed line corresponds to the VMD model
(\ref{fmdF}), while the solid line is for the dipole formula
(\ref{pole}). The averaging
in (\ref{theorformf}) has been performed in the $\pm 3.5\%$ vicinity of the
$\omega$ pole mass. Experimental data are from ref.~\cite{LandPhysRep}.}
\label{fig8}
\end{figure}

In Fig.~\ref{fig8} results of calculations of the extracted FF (\ref{theorformf})
are presented for two different choices  of parametrization
of the input form factors. The dashed line is the
subtracted FF with a VMD parametrization
for both  $\rho$ and $\omega$ mesons,
\begin{equation}
F^{VMD}_{ \omega  \pi^0 {\gamma^*}\,}(s_{\gamma^*}) =-
\dfrac{M_\rho^2}{s_{\gamma^*} - (M_\rho - \frac{i}{2} \Gamma_\rho)^2}; \qquad
F^{VMD}_{ \rho  \,\pi^0 {\gamma^*} }(s_{\gamma^*}) =-
\dfrac{M_\omega^2}{s_{\gamma^*} - (M_\omega - \frac{i}{2} \Gamma_\omega)^2},
\label{fmdF}
\end{equation}
while the solid line is the result for a nonlocal,
pole-like structure of the $\omega$ meson FF
\begin{equation}
F^{pole}_{ \omega  \,\pi^0 \gamma^*}(s_{\gamma^*}) =
\left(1-\frac{s_{\gamma^*}}{(0.65 \, {\rm GeV})^2}\right)^{-1}.
\label{pole}
\end{equation}
For orientation, the previous experimental data on the $\omega$ meson transition
FF, extracted from the reaction
$\pi^- p\to\omega n \to n\pi^0 \mu^+\mu^-$ at pion beam momenta of
25 and 33 $GeV/c$ \cite{LandPhysRep,omegaFrmf1,omegaFrmf2,omegaFrmf3} 
is also presented in Fig.~\ref{fig8}. A
comparison of the extracted FF's with the corresponding inputs
shows that for the considered kinematical conditions, they differ
by less than $0.5\%$ which demonstrates that, if
the cross section is really dominated solely by resonant processes
with $\omega$ and $\rho$ decays alone, the defined ratio
(\ref{theorformf}) can indeed serve as convenient formula to extract the FF's from
experimental data from $pp$ collisions with high accuracy.

However, actually for processes of $NN$ scattering with
a pion and a di-lepton in the final state,
other, non-Dalitz type, diagrams can contribute to the cross section.
In the restricted region  $M_{\pi^0 e^+e^-}\sim M_\omega$ these diagrams
play a role of a smooth background and, in principle, can obscure
the procedure of extracting  FF's by eq.~(\ref{theorformf}). To estimate the
effects of the background one can globally
mimic it by one Feynman diagram with production and decay
 of an effective heavy vector meson into
the considered final state with
effective (freely adjustable) constants.
As seen from eq.~(\ref{cross1}) the structure of the cross section
is as
\begin{equation}
\frac{d\sigma}{ds_{\gamma^*} ds_V}\sim -\int
d\Phi(s_{\gamma^*},s_V,s_{12},P_i) J_\mu^{(NN\to
V)}\left|\frac{M_V^2}{P_V^2-(M_V - \frac{i}{2}\Gamma_V)^2 }\right|^2 J^{+\mu
(NN\to V)} , \label{crossbckgr2}
\end{equation}
where $d\Phi(s_{\gamma^*},s_V,s_{12},P_i)$ is a kinematical function
proportional to the phase space volume $d^5\tau(s_{12},P_V,P_1',P_2')$.
Then it is clear that the resonance structure is governed by the
propagator of the $\omega$ meson, whereas the sub-diagram $NN\to NNV$
provides a smooth dependence of the cross section up on $s_V$.
Correspondingly, one can suppose that the background cross has
the same functional dependence on kinematical variables
as the sub-diagram $NN\to NNV$, i.e.,
it is the same as $J_\mu^{(NN\to V)}(\omega) J^{+\mu (NN\to V)}(\omega)$ with a
non-resonant (constant) propagator. As an easily  trackable procedure
we put the mass of the effective particle  $M_{bkgr}\gg M_\omega$
and adopt the background cross section in the form
\begin{eqnarray}&&
\left(\frac{d\sigma}{ds_{\gamma^*} ds_V}\right)^{(bkgr)} \sim
-\int  \tilde\Phi(s_{\gamma^*},s_V,s_{12},P_i) J_\mu^{(bkgr) }
J^{+\mu ( bkgr)}\, d^5\tau , \label{crossbckgr}
\end{eqnarray}
where $J_\mu^{(bkgr)}\sim \pm J_\mu^{(NN\to NN\omega)}$, and the
function  $ \tilde\Phi(s_{\gamma^*},s_V,s_{12},P_i)$
is chosen such that the contribution of the background
near the $\omega$ pole is $\sim 10 \%$. This order of magnitude
can be estimated from available
experimental data \cite{LandPhysRep}. Note  that the current $J_\mu^{(bkgr)}$, likewise
the $\omega$ and $\rho$ currents, must be transversal, i.e., $J_\mu^{(bkgr)}P_V^\mu=0$,
which implies that this quantity necessarily depends
on kinematical variables, say $J_\mu^{(bkgr)} = J_\mu^{(bkgr)}(s_{\gamma^*},s_V,s_{12},P_i)$.
This means
that this current can not be parameterized  in an arbitrary form§;
at least the condition $J_\mu^{(bkgr)}(s_{\gamma^*},s_V,s_{12},P_i)\ P_V^\mu =0$ must
be fulfilled.

\begin{figure}[h]  %      fig9
\includegraphics[width=1.0\textwidth]{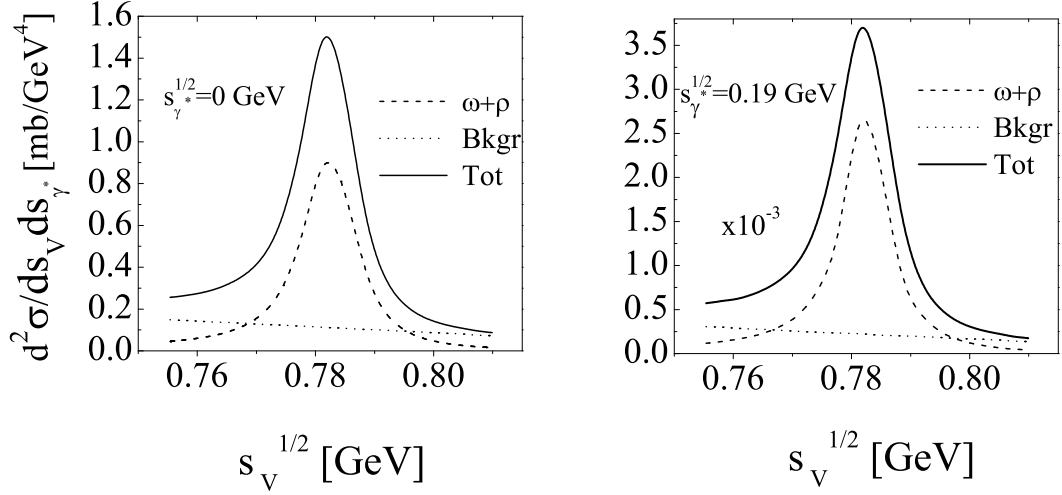} %
\caption{The double differential cross section
$d\sigma/ds_Vds_{\gamma^*}$ as a function of
$s_V^{1/2}$ in the vicinity of the $\omega$ pole for $T_{beam}=2.2\, GeV$
with background contribution taken into account. The relative sign
between the resonant and background amplitudes is chosen positively
resulting in a constructive interference. The dashed line corresponds to the
resonant contributions
of diagrams with $\omega$ and $\rho$ Dalitz decays (cf.\ Fig.~\ref{fig7}),
the dotted line corresponds to the background contribution alone,
and  the solid line is the resulting total cross section. The
employed transition form factor is for the VMD model eq.~(\ref{fmdF}).
Left (right) panel is for $\sqrt{s_{\gamma *}} = 0$ (0.19) GeV.}
\label{fig9}
\end{figure}

\begin{figure}[h]  %      fig10
\includegraphics[width=1.0\textwidth]{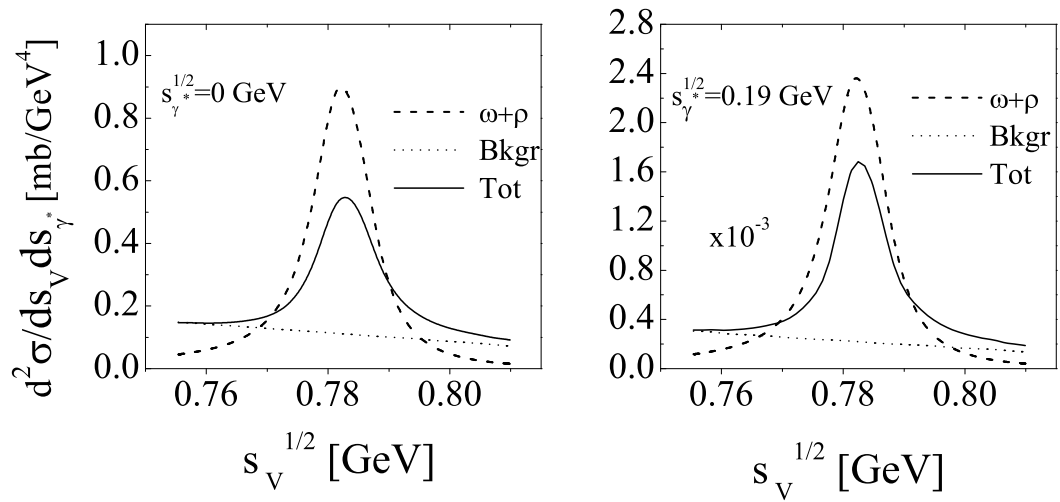} %
\caption{The same as in Fig.~\ref{fig9} but
with a destructive interference of resonant and background contributions.}
\label{fig10}
\end{figure}

\begin{figure}[h]  %      fig11
\includegraphics[width=1.0\textwidth]{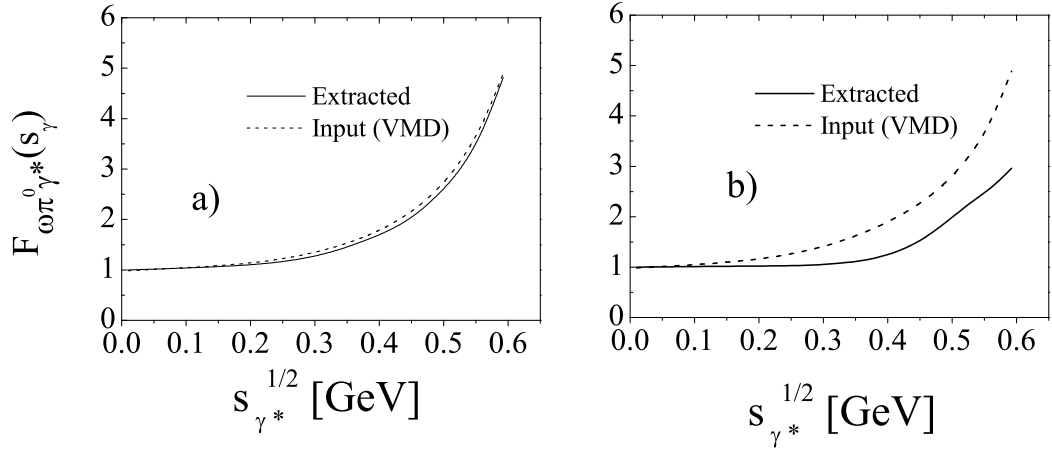} %
\caption{The extracted FF (solid curves)
by using the ratio (\ref{theorformf}) calculated at $T_{beam}=2.2\, GeV$
with inclusion of the background contribution.
Dashed lines correspond to the input FF taken from the
VMD model (\ref{fmdF}). The averaging
has been performed in the $\pm 3.5$ \% vicinity of the
$\omega$ pole mass. Panels a) and b) correspond to destructive and
constructive interference effects, respectively.}
\label{fig11}
\end{figure}

In Figs.~\ref{fig9} and \ref{fig10} results of calculations of the cross section
(\ref{cross1}) with including  the  background (\ref{crossbckgr}) are presented. In
Fig.~\ref{fig9} the relative sign of the background current
is chosen positive (the interference is almost everywhere destructive),
whereas in Fig.~\ref{fig10} the sign is negative
(the interference is mainly constructive). The background
(\ref{crossbckgr}) provides a smooth contribution to the resonant cross section;
at $\omega$  peak it is about $ 10\%$, as dialed. However, the interference effects are rather
important here and can result in  corrections up to $55 \%$ in the maximum.
Figs.~\ref{fig9} and \ref{fig10} also demonstrate that in case of a
constructive interference the resulting cross section (solid lines)
is always larger than the cross section without
background contributions (dashed lines), whereas
in case of a destructive interference the corresponding
cross section is smaller near the peak and larger outside.
These circumstances are rather important in the integrated cross sections
since in the latter case the contribution of the background is partially
compensated in the integral so that the  FF's extracted
via eq.~(\ref{theorformf})  can be quite different in the two cases.
This situation is illustrated
in Fig.~\ref{fig11}, where the input FF and the extracted FF are compared.
One can conclude that a
constructive interference of the background may cause some
uncertainty in the procedure of the experimental determination
of the $\omega$ transition FF's.

\section{Summary}
\label{sumary}

In summary, we have analyzed the di-electron production from Dalitz decays
of light vector mesons produced in $pp$ collisions at intermediate energies.
The corresponding cross section has been
calculated within an effective meson-nucleon approach with
parameters adjusted to describe the free vector
meson production \cite{ourPhi,ourOmega} in nucleon-nucleon reactions near
the threshold. A possible smooth background contribution
to the process has been evaluated as well.
Particular attention is paid to the problem
of whether it is possible to determine in such reactions
the vector meson transition form factors. We argue that by studying
the invariant mass distribution of the final $\pi e^+e^-$ subsystem
as a function of the di-electron mass
in a narrow kinematical interval near the $\omega$ meson mass one can directly  measure
the $\omega$ meson transition form factor $F_{\omega\pi^0\gamma^*}$ in, e.g., $pp$
collisions. Such experiments are
envisaged at HADES and our results may serve as predictions for these
forthcoming experiments. The uncertainties of a procedure to extract 
$F_{\omega\pi^0\gamma^*}$ 
depend upon the scale of the background processes and are expected to be
small if the interference is destructive. Experimental information
on form factors is useful for testing QCD predictions of hadronic quantities
in the non-perturbative domain.

\section*{Acknowledgements}
We thank  H.W. Barz and A.I. Titov for useful discussions.
L.P.K. would like
to thank for the warm hospitality in the Research Center Rossendorf.
This work has been supported by
BMBF grants 06DR121, 06DR135 and the Heisenberg-Landau program.

\appendix

\section{Integration over decay vertices }\label{appA}

The decay part
$\left( \epsilon_{\mu\nu\beta\beta'} P_{\gamma^*}^\mu P_V^{\beta'} j_{\pm}^\nu \right)$
in eq.~(\ref{tok}) can be written in the form of a current-current interaction,
${\cal J}_{(\beta)\nu}(V\to \pi^0\gamma\, ) j_{\pm}^\nu$, where $j_{\pm}^\nu$
is the electromagnetic current of the di-electron, and the decay
current is ${\cal J}_{(\beta)\nu}\sim \epsilon_{\mu\nu\beta\beta'} P_{\gamma^*}^\mu P_V^{\beta'}$.
In the square of the amplitude these currents form the
corresponding electromagnetic ($l^{\nu\nu\prime }$) and
decay (${\cal T}_{(\beta\beta\prime )\nu\nu\prime }$) tensors,
respectively. Obviously, both currents, $j_{\pm}^\nu$ and ${\cal J}_{(\beta)\nu}$, and consequently,
both tensors are conserved:
\begin{equation}
P_{\gamma}^\nu l_{\nu\nu\prime }=
P_{\gamma}^{\nu\prime} l_{\nu\nu\prime }=0; \qquad  P_{V}^\nu {\cal T}_{(\beta\beta\prime)\nu\nu\prime }=
P_{V}^{\nu\prime} {\cal T}_{(\beta\beta\prime)\nu\nu\prime }=0. \label{a1}
\end{equation}
Consider now the integral over the di-electron phase volume.
The electromagnetic tensor (\ref{lmunu})
 depends only on the momenta of the virtual photon and di electron, so that
the   Lorentz structure, after integration over  $R_2(\gamma\to e^+e^-)$,
will  be governed by  only two terms:  one proportional to the metric  tensor $g^{\nu\nu\prime} $
and another one proportional to $P_{\gamma^*}^{\nu}P_{\gamma^*}^{\nu\prime}$:
\begin{eqnarray}
\int l^{\nu\nu\prime}(k_1,k_2,P_{\gamma^*}) R_2(\gamma\to e^+e^-)=
a_1 g^{\nu\nu\prime}+a_2 P_{\gamma^*}^{\nu}P_{\gamma^*}^{\nu\prime}. \label{a2}
\end{eqnarray}
Equation (\ref{a1}) implies that $a_1=-a_2 s_{\gamma^*}$. Multiplying (\ref{a2}) by  $g_{\nu\nu\prime}$
one gets
\begin{eqnarray}&&
a_2=-\frac{1}{3}\int [(k_1P_{\gamma^*}) -s_{\gamma^*}] d\Omega^*_{\bf k}=
\frac{2\pi}{3 },
\label{a3}
\\&&
\int l^{\nu\nu\prime}(k_1,k_2,P_{\gamma^*}) R_2(\gamma\to e^+e^-)=
\frac{2\pi}{3}s_{\gamma^*}\left(-g^{\nu\nu\prime} + \dfrac{P_{\gamma^*}^\nu P_{\gamma^*}^{\nu\prime}}{s_{\gamma^*}}
\right) \label{a4}.
\end{eqnarray}
Analogously, one has  for the decay tensor ${\cal T}_{(\beta\beta\prime)\nu\nu\prime }$
\begin{equation}
\int \epsilon_{\mu\nu\alpha\beta} \epsilon_{\mu\prime\nu\alpha\prime\beta\prime}
P_{\gamma^*}^\mu P_{\gamma^*}^{\mu\prime}   P_V^{\alpha} P_V^{\alpha\prime}
R_2(P_V\to k_\pi+P_{\gamma^*}) = a_1 g^{\beta\beta\prime}+ a_2 P_V^{\beta} P_V^{\beta\prime}
\label{a5}
\end{equation}
with $a_1= -a_2 s_V$ and $a_1=\dfrac{\pi}{12s_V}\lambda^{3/2}(s_V,\mu_\pi^2,s_{\gamma^*})$
and
\begin{equation}
\int \epsilon_{\mu\nu\alpha\beta} \epsilon_{\mu\prime\nu\alpha\prime\beta\prime}
P_{\gamma^*}^\mu P_{\gamma^*}^{\mu\prime}   P_V^{\alpha} P_V^{\alpha\prime}
R_2(P_V\to k_\pi+P_{\gamma^*}) =\dfrac{\pi}{12s_V}\lambda^{3/2}(s_V,\mu_\pi^2,s_{\gamma^*})
\left( -g^{\beta\beta\prime} + \dfrac{P_V^\beta P_V^{\beta\prime}}{s_V} \right).
\label{a6}
\end{equation}

\newpage

%%%%%%%%%%%%%%%%%%%%%%%%%%%%%%%%%%%%%%%%%% BEGIN FIGURES  %%%%%%%%%%%%%%%%%

\end{document}